\documentclass[showpacs,aps,prl,reprint,superscriptaddress,preprintnumbers,nolinenumbers]{revtex4-1}

\usepackage{amsmath}
\usepackage{amssymb}
\usepackage{graphicx}
\usepackage{dcolumn}
\usepackage[dvipsnames]{xcolor}
\usepackage{amsthm,amssymb}
\usepackage{mathrsfs}
\usepackage{color}
\usepackage{endnotes}
\usepackage{bm}
\usepackage{epsfig}
\usepackage{array}
\usepackage{ulem}
\usepackage{color,soul}

\newcommand{\ie}{{\it i.e.\ }}
\newcommand{\CsVSb}{{CsV$_{3}$Sb$_{5}$}}
\newcommand{\RbVSb}{{RbV$_{3}$Sb$_{5}$}}

\newcommand{\Tcdw}{$T_{\rm CDW}$}
\newcommand{\Tc}{$T_{\rm c}$}
\newcommand{\Icsf}{$I_{\rm c,sf}$}

\begin{document}

\title{Giant critical current peak induced by pressure in kagome superconductor RbV$_{3}$Sb$_{5}$}

\author{Lingfei~Wang$^\S$}
\author{Wenyan~Wang$^\S$}
\author{Tsz~Fung~Poon}
\author{Zheyu~Wang}
\author{Chun~Wai~Tsang}
\author{Xinyou~Liu}
\affiliation{Department of Physics, The Chinese University of Hong Kong, Shatin, Hong Kong, China}
\author{Shanmin~Wang}
\affiliation{Department of Physics, Southern University of Science and Technology, Shenzhen, Guangdong, China}
\author{Kwing~To~Lai}
\author{Wei~Zhang}
\affiliation{Department of Physics, The Chinese University of Hong Kong, Shatin, Hong Kong, China}
\author{Jeffery~L.~Tallon}
\affiliation{Robinson Institute, Victoria University of Wellington, Wellington 6140, New Zealand}
\author{Youichi Yamakawa}
\author{Hiroshi Kontani}
\affiliation{Department of Physics, Nagoya University, Furo-cho, Nagoya 464-8602, Japan}
\author{Rina Tazai$^*$}
\affiliation{Yukawa Institute for Theoretical Physics, Kyoto University, Kyoto 606-8502, Japan}
\author{Swee~K.~Goh$^*$}
\affiliation{Department of Physics, The Chinese University of Hong Kong, Shatin, Hong Kong, China}

\date{\today}

\begin{abstract}
Superconductivity can coexist or compete with other orders such as magnetism or density waves. Optimizing superconductivity requires identifying competing orders that may disrupt Cooper pair coherence. Here, we use the self-field critical current ($I_{\rm c,sf}$) to probe pressure-tuned superconductivity in the kagome superconductor RbV$_3$Sb$_5$. As pressure destabilizes the charge-density wave (CDW) state, $I_{\rm c,sf}$ drastically enhances, peaking near the critical pressure where the CDW state is completely suppressed at zero temperature. Surprisingly, a weaker $I_{\rm c,sf}$ peak emerges within the CDW phase. Near the pressure of the weaker peak, the superconducting phase transition temperature shifts from an increasing trend with pressure to a near plateau. Our analysis suggests the possibility of a sudden change in the CDW pattern or a Lifshitz transition, highlighting the need for microscopic examinations of the CDW state for understanding the pressure evolution of superconductivity in RbV$_3$Sb$_5$.

\end{abstract}

\maketitle

The kagome lattice has emerged as a versatile platform to host exotic quantum states due to its unique geometry. Composed of corner-sharing triangles, the kagome lattice introduces geometrical frustration, which has long been a desirable configuration for quantum spin liquids~\cite{bergman2008,balents2010,norman2016,Zhou2017}. When electrons are mobile on a kagome lattice, a kagome metal is formed. Kagome metals are particularly fascinating because they combine geometrical frustration and itinerant electrons, leading to rich physics~\cite{Kiesel2013, andreanov2020,tazai2022,wu2022,Kang2023,han2023,yin2022}. The kagome bandstructure inherently exhibits flat bands and Dirac cones~\cite{wang2013,Kiesel2012,wen2010,hu2022}. Consequently, the interplay between strong electronic correlations and topology can be explored by tuning the Fermi energy~\cite{li2022,Oey2022,Du2022,wenzel2023,Du2021,Chen2021a,Wang2021}. In short, kagome metals are fascinating tunable systems in quantum materials research.

\RbVSb\ is a member of the AV$_3$Sb$_5$ (A = K, Rb, Cs) family where the vanadium atoms form a perfect kagome net. At ambient pressure, \RbVSb\ undergoes a transition into a charge-density wave (CDW) state below $T_{\rm CDW}=102~$K~\cite{Ortiz2019,Cho2021,Li2021,Yin2021,Liu2021,wu2022}. Below approximately 0.92~K, \RbVSb\ exhibits superconductivity, making it one of the rare kagome metals with a superconducting ground state. The CDW state in \RbVSb\ is associated with a range of interesting phenomena, including the giant anomalous Hall effect and the possibility of a time-reversal symmetry broken state~\cite{xu2022,guguchia2023,wang2023,wang2023AHE}.  The superconducting state has also been proposed to have an unconventional origin in some reports~\cite{guguchia2023,frassineti2023,wu2021,Wang2021a}. Both \Tcdw\ and the superconducting transition temperature (\Tc) are tunable by pressure -- under pressure, \Tcdw\ decreases monotonically and extrapolates to 0~K at $p_{\rm c}\approx23$~kbar, while \Tc\ initially increases below $p'\approx13$~kbar, then shows a rather weak pressure dependence from $p'$ to just above $p_{\rm c}$~\cite{Du2022,guguchia2023,Wang2021a}. The pressure dependence of \Tc\ is noticeably different from the results of the sister compound \CsVSb, where \Tc\ exhibits a double-dome behavior within the CDW region~\cite{Liang2021,kang2022,Chen2021a}. Hence, it is important to investigate the unusual competition between superconductivity and CDW in \RbVSb\ to find out why \Tc\ ceases to increase above $p'$.

One approach to understand the sudden change in the pressure dependence of \Tc\ at $p'$ is to traverse the temperature-pressure ($T$-$p$) phase diagram at zero temperature to identify anomalous behavior. Here, we measure the self-field transport critical current ($I_{\rm c,sf}$), \ie\ transport critical current without applying an external magnetic field, from ambient pressure to pressures beyond $p_{\rm c}$. It has been demonstrated that \Icsf(0), namely \Icsf\ at the zero-temperature limit, is sensitive to quantum phase transitions. In \CsVSb, a peak in \Icsf(0) is detected near the pressure at which the CDW order is fully suppressed~\cite{Wang2024}. Similar behavior has also been reported in cuprates~\cite{naamneh2014,talantsev2015}, heavy fermions~\cite{jung2018} and quasi-skutterudites~\cite{liu2022}.
Surprisingly, our measurements uncover two peaks in the pressure dependence of \Icsf, a pronounced peak centered at $p_{\rm c}$, and a weaker peak at $p'$. Thus, our work suggests the existence of two quantum phase transitions, one of them yet to be fully established, which underpin the $T$-$p$ phase diagram of \RbVSb.
\\

\begin{figure}[!t]
      \hspace*{-0.3cm}
      \resizebox{9.6cm}{!}{
              \includegraphics{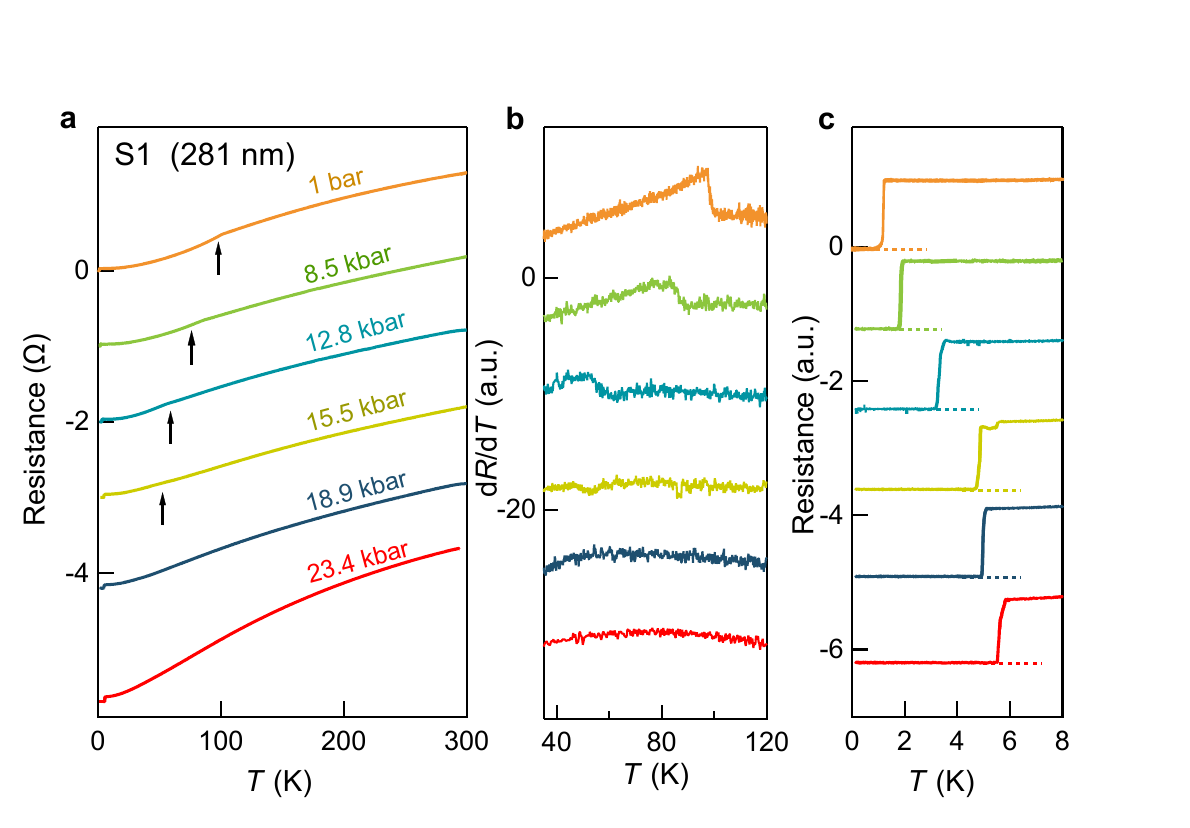}}              
              \caption{\label{fig1} 
              a) Temperature dependence of the electrical resistance $R$ at various pressures in \RbVSb\ (S1). The black arrows indicate the charge density wave transition temperature $T_{\rm CDW}$. The traces are vertically offset for clarity. b) Corresponding temperature derivatives $dR/dT$. The CDW transition (sharp peak feature)  becomes indistinguishable above 18.9~kbar.
              c) Normalized superconducting transitions shown in $0–8$~K range. The dashed lines extrapolate to zero resistance, and  $T_c$ is defined as the temperature where the resistance reaches zero and is used for the calculation of 2$\Delta_{0}/k_{\rm B}T_{\rm c}$ in Figure~\ref{fig4}.       }   
\end{figure}

\noindent {\bf \large Results and Discussion}\\
\noindent {\bf Evolution of the Electronic States under Pressure}\\
Electrical transport gives prominent signatures at \Tcdw\ and \Tc, offering a straightforward means to track their pressure dependence. We prepared three device-integrated diamond anvil cells (DIDAC), each equipped with multiple prepatterned electrodes, to measure thin flakes with thicknesses of 281~nm (S1), 170~nm (S2) and 131~nm (S3). See Methods for experimental details. Figure~\ref{fig1}a displays representative temperature dependence of resistance $R(T)$ curves for S1. At ambient pressure, a kink appears in $R(T)$ at 98.6~K, as indicated by the vertical arrow in Figure~\ref{fig1}a. This corresponds to the well-documented CDW transition, and the transition temperature (\Tcdw) is clearly visible in the first derivative (Figure~\ref{fig1}b). At low temperatures, $R(T)$ suddenly drops to zero, signifying a superconducting transition (Figure~\ref{fig1}c).

Our data on S1 confirm the reported pressure dependence of both \Tcdw\ and \Tc. As displayed in Figure~\ref{fig1}, when pressure increases, \Tcdw\ decreases while \Tc\ increases. Data collected using S2 and S3 behave in a similar manner, and are presented in Supporting Information. Combining \Tcdw\ and \Tc\ under pressure from all samples, we construct the $T$-$p$ phase diagram depicted in Figure~\ref{fig4}, which will be discussed in greater detail later.  

~\\

\begin{figure*}
    \centering
    \setlength{\abovecaptionskip}{0cm}
    \resizebox{16cm}{!}{
        \includegraphics[width=1\textwidth]{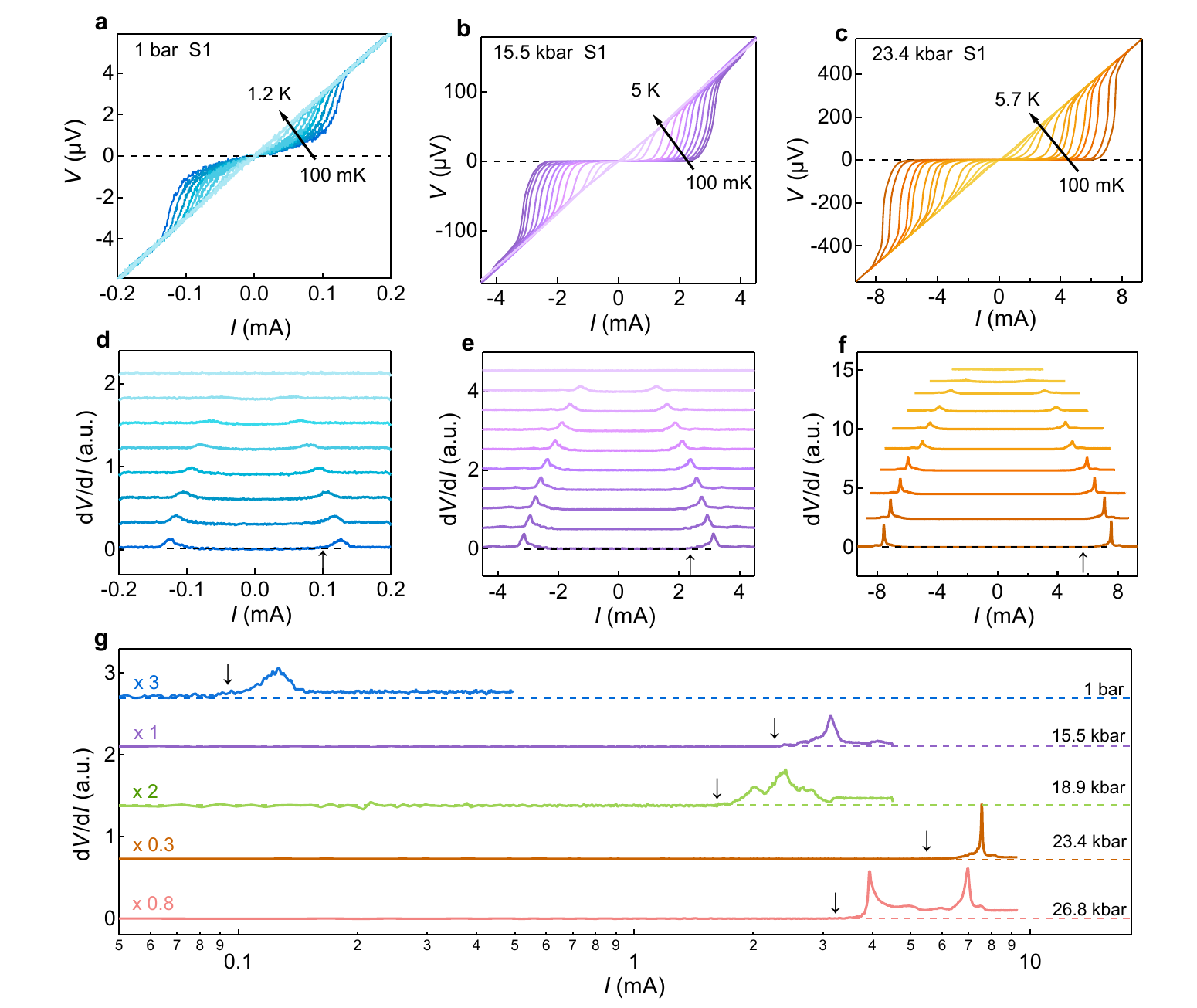}} \caption{\label{fig2} 
        {a) $V$-$I$ characteristics for sample S1 at 1~bar, b) 15.5~kbar and c) 23.4~kbar. The calculated first derivative of $V(I)$, $dV/dI$ for each pressure are shown in d, e) and f), respectively. g) $dV/dI$ curves at 100~mK for five selected pressures, plotted with a logarithmic scale on the $x$-axis. Numerical labels adjacent to curves indicate multiplicative scaling factors applied to the ordinate ($y$-axis) values for clarity.  Short arrows in (d-g) denote the onset current ($I_c$), defined as the deviation point from $dV/dI = 0$. $V$-$I$ characteristics and the associated first derivatives for all other pressures are presented in Supporting Information.}}
\end{figure*}

\medskip \noindent {\bf $I_{\rm c,sf}$ as a Probe of the Superconducting State}\\
To probe the superconducting state, we collected a series of low-temperature voltage-current ($V$-$I$) characteristics at various pressures. Figure~\ref{fig2}a-c shows the $V$-$I$ curves of S1 at 1~bar, 15.5~kbar and 23.4~kbar. At the lowest temperature, a rapid increase in voltage beyond a threshold current signifies a transition from the superconducting state to the normal state, as can be seen in all three datasets. At 1~bar, the threshold current is much smaller, and the $V$-$I$ curve exhibits a small but finite slope in the superconducting state. This finite slope is an experimental artifact, corresponding to a resistance offset in the superconducting state. However, the offset does not affect the determination of \Icsf. Following our earlier works~\cite{zhang2023, Wang2024}, $I_{\rm c,sf}$ is defined as the current at which $dV/dI$ deviates from zero. In Figure~\ref{fig2}d-f, we plot $dV/dI$ corresponding to the $V$-$I$ data in Figure~\ref{fig2}a-c, and the short vertical arrows denote \Icsf. As the temperature rises, \Icsf\ decreases, consistent with the expected behavior.

From Figure~\ref{fig2}a-f, it is evident that \Icsf\ exhibits a strong pressure dependence. To facilitate the comparison, we plot $dV/dI$ at 100~mK for S1 at five selected pressures, as illustrated in Figure~\ref{fig2}g. Two important observations on \Icsf\ at the zero-temperature limit can be readily made from Figure~\ref{fig2}g. First, \Icsf\ increases by nearly two orders of magnitude when pressure increases from 1~bar to 23.4~kbar (note the logarithmic scale for the current axis). Second, \Icsf\ does not vary monotonically with pressure. At 18.9~kbar \Icsf\ is smaller than that at 15.5~kbar. Similarly, \Icsf\ shows a peak near 23.4~kbar. Thus, two \Icsf\ peaks are observed within the entire pressure range covered.

\begin{figure*}[!t]
        \centering
       \resizebox{16cm}{!}{
              \includegraphics{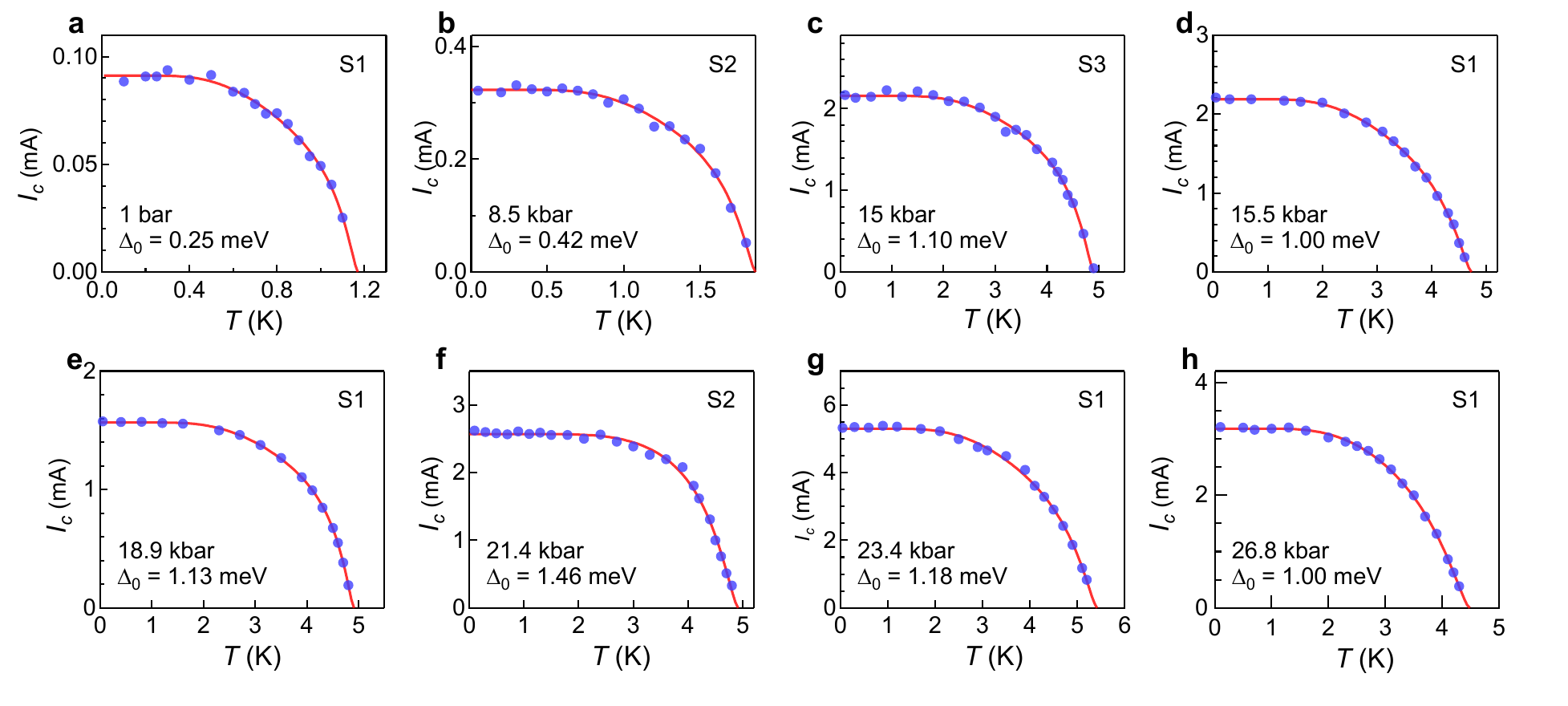}}                	
              \caption{\label{fig3}            
              {
              Temperature Dependence of \Icsf\ under Various Pressures. 
              Experimentally obtained $I_{\rm c,sf}(T)$ at eight representative pressures. Sample classification: S1 (a,d,e,g,h), S2 (b,f), S3 (c). The solid curves are the single $s$-wave gap fits. The extracted superconducting gap $\Delta_0$ are 0.25~meV at 1~bar, 0.42~meV at 8.5~kbar, 1.10~meV at 15~kbar, 1.00~meV at 15.5~kbar, 1.13~meV at 18.9~kbar, 1.46~meV at 21.4~kbar, 1.18~meV at 23.4~kbar and 1.00~meV at 26.8~kbar, respectively.  See Supporting Information for $I_{\rm c,sf}(T)$ at all other pressures. }}
\end{figure*}

\begin{figure}[!t]\centering
      \resizebox{10cm}{!}{
 \includegraphics{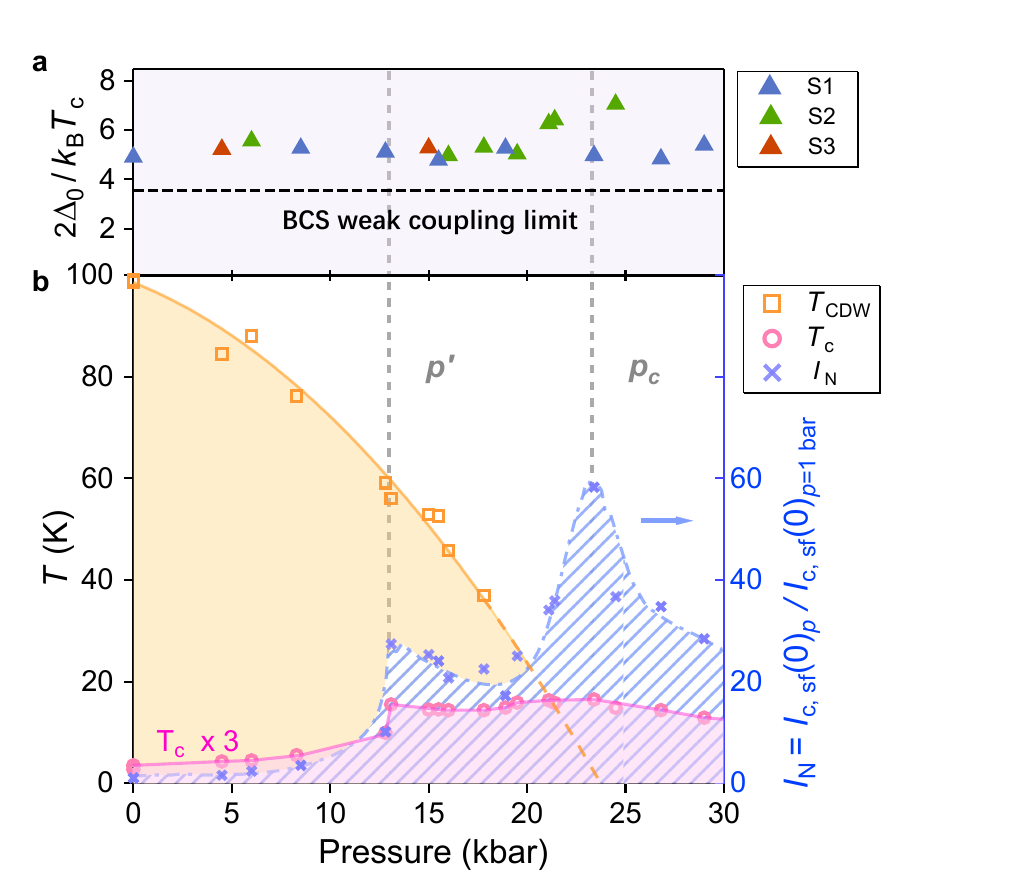}}       
 \caption{\label{fig4} a) Ratio of superconducting gap to critical temperature ($2\Delta_0/k_BT_c$) at various pressure in three samples. The black dash line indicates the BCS weak-coupling limit. b) Temperature-pressure phase diagram.  \Tcdw\ and \Tc\ correspond to orange and pink hollow markers, respectively. Blue cross symbols denote the critical current ratio $I_{\rm N}$~= $I_{\rm c,sf}$(0)$_p$ / $I_{\rm c,sf}$(0)$_{p = 1 \rm bar}$. For clarity, \Tc\ values are scaled by a factor of three. The gray dashed lines demarcate  $p'$ and $p_c$.}     
\end{figure}

Next, we investigate the temperature dependence of \Icsf\ in detail. Figure~\ref{fig3} shows $I_{\rm c,sf}(T)$ at various pressures. At all applied pressures, $I_{\rm c,sf}(T)$ exhibits analogous behavior: upon cooling from $T_{\rm c}$, it undergoes a rapid increase before saturating at sufficiently low temperatures. This characteristic behavior is qualitatively consistent across all pressure regimes and the three samples studied. Extensive studies~\cite{talantsev2015,talantsev2017,zhang2023,liu2022,semenok2022} demonstrate that the zero-field transport critical current density $J_{\rm c,sf}$ correlates directly with the superfluid density $\rho_{\rm s} \propto \lambda^{-2}$, where $\lambda$ is the London penetration depth. Crucially, for thin flake samples with half-thickness $b$ satisfying $b \ll \lambda(0)$, where $\lambda(0)$ is the penetration depth at the zero-temperature limit, $J_{\rm c,sf}$ conforms to the relation derived in~\cite{talantsev2015,talantsev2017}.

\begin{equation}
J_{\rm c,sf}=\frac{\phi_0}{4\pi\mu_0\lambda^3}\left(\ln\left(\frac{\lambda}{\xi}\right)+0.5\right)
\label{eqn_thin}
\end{equation}
where $\phi_0$ is the flux quantum, $\mu_0$ is the vacuum permeability, and $\xi$ is the coherence length.

The measured thicknesses $2b$ for samples S1, S2, and S3 are 281~nm, 170~nm, and 131~nm, respectively. The 0~K limit penetration depth $\lambda(0) \approx 690$~nm, as determined by $\mu$SR measurements~\cite{guguchia2023}, ensures that the half-thickness $b$ of all samples is significantly smaller than $\lambda(0)$, satisfying the criterion $b \ll \lambda(0)$. Consequently, the application of the thin-limit expression (Eq.~(\ref{eqn_thin})) is validated. Since the logarithmic term $\ln(\lambda/\xi)$ exhibits a negligible temperature variation, the dominant temperature dependence of $J_{\rm c,sf}(T)$ arises from $\lambda^{-2}(T)$. This implies that the normalized ratio $[J_{\rm c,sf}(T)/J_{\rm c,sf}(0)]$, identical to [$I_{\rm c,sf}(T)/I_{\rm c,sf}(0)$] because of the cancellation of the sample geometry factor, scales as $[\lambda^{-2}(T)/\lambda^{-2}(0)]^{3/2}$, which is governed by the form of the superconducting gap.

In \CsVSb, multiple experiments consistently demonstrate the nodeless $s$-wave superconductivity, including the detection of a Hebel-Slichter coherence peak below $T_{\rm c}$~\cite{mu2021s} by nuclear magnetic resonance, and penetration depth results from muon spin rotation ($\mu$SR) and tunnel diode oscillator (TDO) ~\cite{gupta2022microscopic,gupta2022two,shan2022muon,duan2021nodeless}. Furthermore, self-field critical current analyses also confirmed the nodeless nature of superconductivity at ambient pressure as well as under high pressures~\cite{zhang2023,Wang2024}. 
 
Motivated by these studies, we first try the simplest gap model, namely $\lambda^{-2}(T)/\lambda^{-2}(0)$ governed by an $s$-wave superconducting gap: 
\begin{equation}
\frac{\lambda^{-2}(T)}{\lambda^{-2}(0)}=1-2\sqrt{\frac{\pi\Delta_0}{k_{\rm B}T}}\exp\left(-\frac{\Delta_0}{k_{\rm B}T}\right).
\label{swave}
\end{equation}
As demonstrated in Figure~\ref{fig3}, the combination of Eq.~(\ref{eqn_thin}) and Eq.~(\ref{swave}) provides an excellent description of the experimental data at all measured pressures. Hence, a simple $s$-wave gap is a sufficient description of \Icsf\ data.  These analyses enable the extraction of the pressure dependence of the superconducting gap at the zero-temperature limit, $\Delta_0$. At ambient pressure, $\Delta_0$ = 0.25 meV, which corresponds to 2.46 $k_{\rm B}T_{\rm c}$. When $T_{\rm c}$ is appreciably higher, $\Delta_0$ is also higher. In fact, the ratio $2\Delta_0/k_{\rm B}T_{\rm c}$ is rather constant throughout the pressure range, but it is elevated above the Bardeen–Cooper–Schrieffer (BCS)  weak-coupling limit of 3.53. These results demonstrate a pressure-invariant strong-coupling behavior persisting across both the coexistence and the pure superconducting phases. 
This persistent strong-coupling superconductivity across the pressure-tuned CDW phase boundary emphasizes the robustness of the underlying pairing mechanism in \RbVSb. Significantly, an enhanced \Icsf\ accompanies this superconductivity near the critical pressures. Further details will be discussed in the next section.

~\\

\noindent {\bf Drastic Enhancement and Double-peak Feature of $I_{\rm c,sf}$ under Pressure}\\
Combining the datasets from the three preceding figures, we construct the $T$-$p$ phase diagram in Figure~\ref{fig4}. Our phase diagram is consistent with that reported by Du {\it et al.}~\cite{Du2022}. Additionally, Figure~\ref{fig4} displays the pressure dependence of the zero-temperature self-field critical current \Icsf(0), providing a clear view of the drastic enhancement and the double-peak feature of $I_{\rm c,sf}$(0) on the backdrop of the pressure-tuned phase boundaries. Following the procedure established in our previous work~\cite{Wang2024}, we normalize \Icsf(0) under pressure (\Icsf(0)$_p$) to the ambient pressure value \Icsf(0)$_{p = 1 \rm bar}$ for each sample to enable the comparison across different samples. The normalized critical current is defined as $I_{\rm N}$~= $I_{\rm c,sf}$(0)$_p$ / $I_{\rm c,sf}$(0)$_{p = 1 \rm bar}$. Notably, $T_{\rm c}$ reaches a maximum of 5.2~K and saturates above $p' \approx 13$~kbar, where the first \Icsf(0) peak emerges with a $\sim\!27$-fold enhancement (i.e. $I_{\rm N}\approx 27$) relative to ambient pressure. This is followed by a more dramatic peak at $p_{\rm c} \approx 23$~kbar, where $T_{\rm CDW}$ extrapolates to 0~K, with $I_{\rm N}$ reaching as high as $\sim$60. Such a large critical current enhancement is extraordinary - in \CsVSb, $I_{\rm N}$ is only $\sim$10 at the corresponding $p_c$~\cite{Wang2024}.

In general, when $T_{\rm c}$ is enhanced, the critical current might be expected to be enhanced. However, such a scenario is not applicable here. To demonstrate this, we simply observe that $T_{\rm c}$ is relatively pressure-independent around $p_{\rm c}$ while $I_{\rm N}$ exhibits a peak there. To be more quantitative, we plot $I_{\rm N}/T_{\rm c}$ and $I_{\rm N}/T_{\rm c}^{1.5}$ against pressure (see Supporting Information). The characteristic double-peak structure persists in both presentations, ruling out the possibility that $I_{\rm N}$ is solely governed by the variation of $T_{\rm c}$. Note that the expectation $I_{\rm N}\propto T_{\rm c}^{1.5}$ comes from the combination of Eq.~(\ref{eqn_thin}) and the Uemura relation~\cite{Wang2024}. Thus, the double-peak feature of $I_{\rm N}$ is new and genuine, and as we shall discuss below, this pressure dependence of $I_{\rm N}(p)$ underpins the extraordinary pressure dependence of $T_{\rm c}$ that distinguishes it from that observed in the more extensively studied \CsVSb.

~\\

\noindent {\bf Origin of Double-peak Feature}\\
We have revealed a remarkable double-peak feature in the pressure dependence of \Icsf(0) in kagome superconductor \RbVSb. The large peak at $p_{\rm c} \approx 23$ kbar is naturally associated with the suppression of charge order, given that this peak is located near the border of the charge order. In fact, similar critical current behaviors have been observed in several quantum critical systems. For instance, in the cuprate superconductor YBa$_2$Cu$_3$O$_{7-\delta}$, $I_{\rm c,sf}$(0) exhibits a peak at an optimal hole doping level $p \approx$ 0.19, where the pseudogap closes~\cite{talantsev2015}. Similarly, a significant enhancement of $I_{\rm c,sf}$(0) occurs at the critical pressure where the antiferromagnetic transition temperature extrapolates to 0 K in heavy fermion CeRhIn$_5$~\cite{jung2018}. In our previous work, the critical current density $J_{\rm c,sf}$(0) peaks near a structural QCP in (Ca$_x$Sr$_{1-x}$)$_3$Rh$_4$Sn$_{13}$~\cite{liu2022}. In another vanadium-based kagome superconductor CsV$_3$Sb$_5$, a peak in $I_{\rm c,sf}$(0) is also observed when the CDW phase is fully suppressed~\cite{Wang2024}. Therefore, the stronger $I_{\rm c,sf}$(0) peak detected in \RbVSb\ at $p_{\rm c}$ could originate from enhanced quantum fluctuations due to the weakening of the CDW order. 
 
\begin{figure}[!t]\centering
      \resizebox{8.5cm}{!}{
 \includegraphics{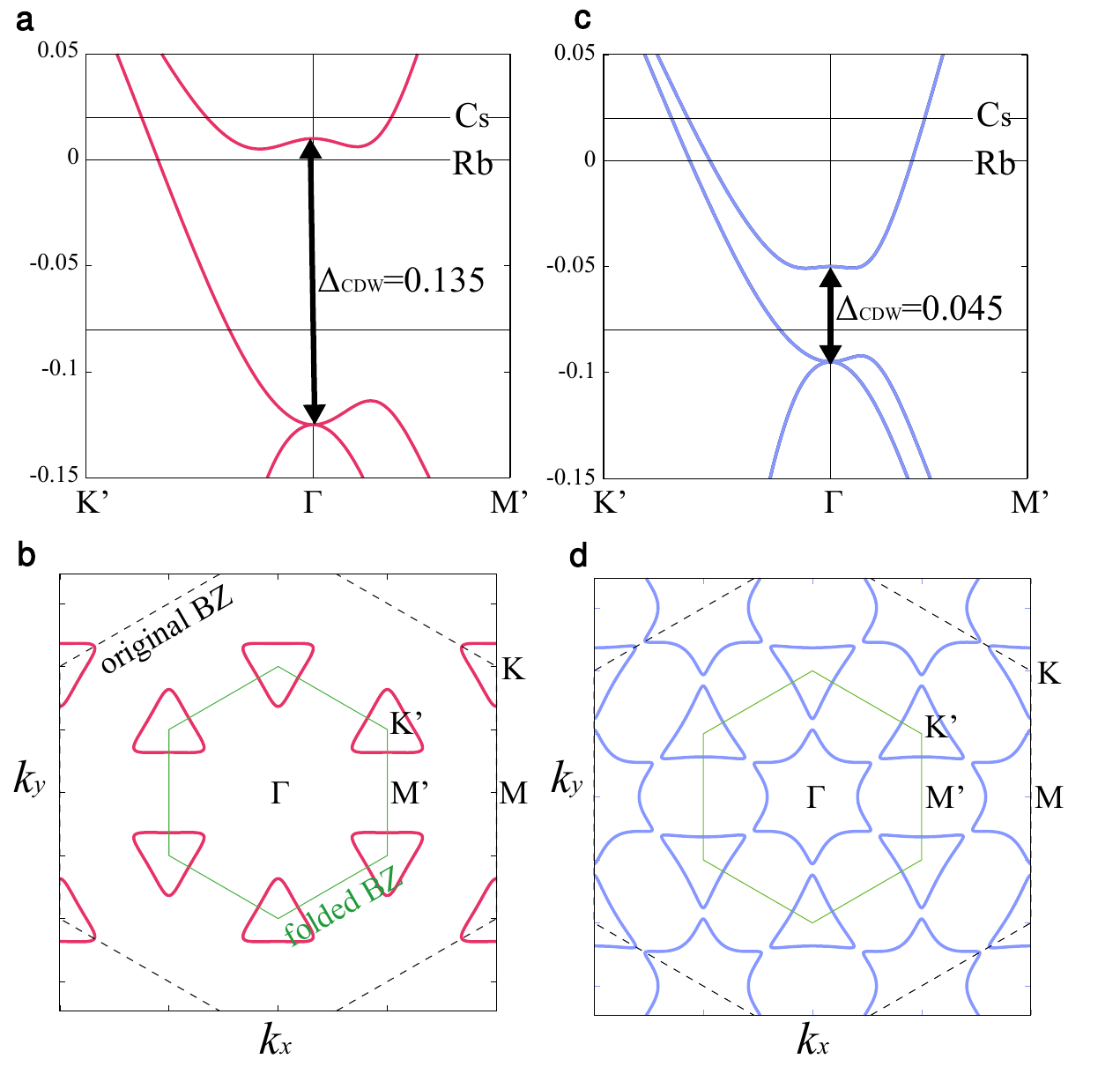}}       
 \caption{\label{fig5} a) The band dispersion and b) Fermi surface at $\Delta_{\rm{CDW}} =0.135$~eV.
 c, d) The corresponding band dispersion and Fermi surface at a higher pressure with a smaller $\Delta_{\rm{CDW}}$ of $0.045$~eV. The horizontal lines in (a,b) indicate, from top to bottom, the Fermi energy of \CsVSb, the Fermi energy of \RbVSb, and the energy of the (degenerate) van Hove singularities.}    
\end{figure}

Although the origin of the \Icsf\ peak at $p_{\rm c}$ is now understood, the formation of the peak within the CDW phase (at $p' \approx 13$~kbar) is surprising. Notably, this \Icsf\ anomaly emerges at the pressure where the pressure dependence of $T_{\rm c}$ changes abruptly: from a strong increase below $p'$ to a weakened pressure dependence above it. Keeping in mind that \Icsf\ is sensitive to zero-temperature anomalies, the appearance of the peak at $p'$ suggests a possible change in the ground state which eventually influences superconductivity. Some scenarios include the restructuring of the CDW state associated with the bond order fluctuations or a modification of the superconducting gap~\cite{tazai2022,tazai2023}. A pressure-induced switch from nodal to nodeless superconductivity has been reported near $p'$~\cite{guguchia2023}. However, our \Icsf\ data do not support this scenario (Figure~\ref{fig3}), and a recent penetration depth measurement indicates a nodeless gap at ambient pressure~\cite{nagashima2025}.  Furthermore, the pressure-independent behavior of $2\Delta_0/k_{\rm B}T_{\rm c}$ across $p'$ indicates that a change of pairing symmetry is unlikely to occur.  

Instead, a sudden change in the CDW ordering vector \textbf{Q} at $p'$ is a possible explanation. This has been reported in the sister compound \CsVSb, where the CDW state changes from triple-Q to a stripelike structure~\cite{Zheng2022}. In general, the V atoms within a given kagome layer can form either Star-of-David (SoD) or trihexagonal (TrH) patterns. Furthermore, when these kagome layers are stacked along the $c$-axis, various stacking patterns can occur. For example, there could be TrH configurations $\pi$ shifted between the adjacent kagome layers, or a mixture of TrH and SoD, giving rise to additional periodicity along the $c$-axis. Under pressure, it is conceivable that the kagome layers in the CDW state adopt a configuration different from the ambient pressure case. Future experiments, such as nuclear magnetic resonance or diffraction experiments under pressure, will shed light on this possibility.

Another possibility that explains the pressure-induced jump in $T_{\rm c}$ is a Lifshitz transition. 
As pressure is applied, the CDW transition temperature decreases gradually, and the corresponding CDW gap is expected to decrease.
It can be demonstrated theoretically that this change can induce a Lifshitz transition.
In Figure~\ref{fig5}, we plot the band dispersions and Fermi surface in the TrH CDW state based on the single $d_{xz}$-orbital model ~\cite{nagashima2025}. The model is given as 
 $t_{ij}+\delta t_{ij}$, where
$t_{ij}$ is the original hopping with $t=-0.5~\rm{eV}$ for the nearest neighbor sites and $t'=-0.08~\rm{eV}$ for the third nearest ones.
$\delta t_{ij}$ is the modulation of the hopping in the nearest neighbor sites due to the $2\times 2$ TrH CDW.
To simplify the discussion, we only adopt the folded Brillouin zone due to the $2 \times 2$ CDW for the following comparisons. Without the CDW, the three van hove singularities at $E_{\rm {vHS}}$ are degenerate at the $\Gamma$ point in the folded Brillouin zone. 
When the CDW appears,
the gap opens with a magnitude
$\Delta_{\rm{CDW}}=6 \delta t$.
Figure~\ref{fig5}b,d show
the Fermi surface in the TrH CDW state for two illustrative gap values: $\Delta_{\rm{CDW}}=0.135~\rm{eV}$ for the ambient pressure case, and a smaller gap $\Delta_{\rm{CDW}}=0.045~\rm{eV}$ for the higher pressure scenario.
When the gap is small, an electron pocket around the $\Gamma$-point appears; as the gap becomes larger, the pocket disappears, indicating a Lifshitz transition.
Concomitantly, the density of states changes, which could account for the jump in $T_{\rm c}$ at $p'\approx13$~kbar. 
Compared with \CsVSb\ ($E_{\rm {vHS}}\sim -0.1~\rm{eV}$), the van Hove singularity in \RbVSb\ ($E_{\rm {vHS}}\sim -0.08~\rm{eV}$) is located closer to the Fermi energy, and therefore such a transition would naturally be observed only in \RbVSb, consistent with our experiment.
Importantly, the proposed Lifshitz transition scenario is well justified in \RbVSb\, since the $d_{xz}$ orbitals give rise to nearly ideal two-dimensional Fermi surfaces. To verify the possibility of a Lifshitz transition, detailed magnetotransport studies near $p'$ will be instructive.
\newline

\noindent {\bf \large  Conclusion}\\
In summary, our pressure-dependent study of the self-field critical current \Icsf\ in the kagome superconductor \RbVSb\ reveals a striking double-peak feature in the zero-temperature limit. A prominent peak occurs near the critical pressure of $p_{\rm c}\approx23$~kbar, where the CDW state is completely suppressed. This peak can be attributed to enhanced quantum fluctuations. Surprisingly, a distinct, weaker peak appears at $p'\approx13$~kbar within the CDW phase. This weaker peak coincides precisely with a significant change in the pressure dependence of the $T_{\rm c}$ $-$ from a strong initial increase to a plateau. We propose that pressure induces a Lifshitz transition at $p'$, or a reconstruction of the CDW order, both can affect the $T_c$ without changing the robust nodeless, strong-coupling pairing symmetry indicated by our \Icsf\ analyses. These findings reveal a complex interplay of competing orders and underscore the necessity to identify the nature of the `hidden' quantum phase transition within the CDW state for a comprehensive understanding of the superconductivity in \RbVSb.

~\\


\noindent {\bf \large Experimental Section}\\
\it Sample Synthesis, Characterization and Exfoliation: \rm High-quality single crystals of \RbVSb\ were synthesized from Rb (ingot, 99.95 $\%$), V (powder, 99.9 $\%$) and Sb (shot, 99.9999 $\%$) using the self-flux method similar to Ref.~\cite{Yin2021}. The as-grown single crystals were millimeter-sized shiny plates. Three \RbVSb\ single-crystal specimens (designated S1, S2, and S3) from the same synthesis batch were mechanically exfoliated using blue Nitto tape and silicone elastomer polydimethylglyoxime (PDMS). Selected flakes were subsequently transferred onto diamond substrates with pre-patterned gold electrodes for high-pressure measurements. X-ray diffraction (XRD) measurements were performed on a Rigaku diffractometer utilizing Cu$K_\alpha$ radiation at room temperature. Chemical composition analysis was conducted via an Oxford energy-dispersive X-ray spectroscopy (EDX) integrated with a JEOL JSM-7800F scanning electron microscope. Sample thicknesses were determined using a Thermo Scientific Scios 2 DualBeam FIB system, with S1 281~nm, S2 170~nm and S3 131~nm.

\it Diamond Anvil Cells: \rm Electrical transport measurements on \RbVSb\ thin flakes under high pressure were performed using the device-integrated diamond anvil cell (DIDAC) technique~\cite{Wang2024,Xie2021,zhang2023}. To establish electrical contacts, a six-probe Au microelectrode pattern was first fabricated on the diamond culet (800~$\mu$m and 1000~$\mu$m in diameter) employing photolithography and physical vapor deposition (PVD). An insulating layer consisting of alumina/Stycast-1266 mixture was then applied between the stainless steel gasket and the microelectrodes. Subsequently, the exfoliated \RbVSb\ thin flake was transferred onto the center of the culet. Finally, to prevent oxidation, the \RbVSb\ flake was covered by a thin film of h-BN.

\it Electrical Transport and Critical Current Measurements: \rm Electrical resistance measurements were conducted using a standard four-terminal configuration integrated within a Physical Property Measurement System (PPMS) by Quantum Design. Measurements of the self-field critical current and superconducting transition were performed within a Bluefors dilution refrigerator. Stable electrical contacts to the \RbVSb\ thin flake were established using the six-probe Au microelectrodes described previously. We take the average of charge density wave transition temperatures from cool-down and warm-up processes as \Tcdw. The voltage-current ($V$-$I$) characteristics were acquired using a Keithley 2182A nanovoltmeter coupled with a Keithley 6221 current source operated under pulsed delta mode. Pulsed excitation mode was implemented to minimize Joule heating, with pulse durations of 11 ms and inter-pulse intervals of 1 s.

~\\
~\\

\noindent {\bf \large Acknowledgements}\\
\noindent The work was supported by the Research Grants Council of Hong Kong (CUHK 14300722, CUHK 14301020, CUHK 14301725 and CUHK 14002724), CUHK Direct Grant (4053577, 4053664), JSPS KAKENHI (JP25H01246, JP25H01248, JP22K14003, JP24K00568, JP24K06938), and the Guangdong Provincial Quantum Science Strategic Initiative (GDZX2301009).

~\\

\noindent {\bf \large Author contributions}\\
$^*$tazai.rina.2y@kyoto-u.ac.jp, skgoh@cuhk.edu.hk
\\

\noindent L.W. and W.W. contributed equally to this work. S.K.G. conceived the project. L.W. and W.W. led the experiments and analyzed data. Z.W., X.L. and W.Z. assisted in pressure cells preparation and methodology development. T.F.P., C.W.T. and K.T.L. synthesized the single crystals. S.W. provided hBN. J.L.T., Y.Y., H.K. and R.T. provided theoretical support. S.K.G., L.W., W.W. and R.T. wrote the paper with input from all authors. 

~\\

\noindent {\bf \large Data Availability Statement}\\
\noindent The data that support the findings of this study are available from the corresponding authors upon reasonable request.

~\\

\noindent {\bf  \large Keywords}\\
\noindent kagome superconductors, quantum phase transition, critical current, high pressure





\providecommand{\noopsort}[1]{}\providecommand{\singleletter}[1]{#1}%

\end{document}